\documentclass[lineno]{biometrika}

\usepackage{amsmath}

\usepackage{times}
\usepackage{bm}
\usepackage{natbib}

\usepackage[plain,noend]{algorithm2e}

\makeatletter
\renewcommand{\algocf@captiontext}[2]{#1\algocf@typo. \AlCapFnt{}#2} 
\renewcommand{\AlTitleFnt}[1]{#1\unskip}
\def\@algocf@capt@plain{top}
\renewcommand{\algocf@makecaption}[2]{%
  \addtolength{\hsize}{\algomargin}%
  \sbox\@tempboxa{\algocf@captiontext{#1}{#2}}%
  \ifdim\wd\@tempboxa >\hsize
    \hskip .5\algomargin%
    \parbox[t]{\hsize}{\algocf@captiontext{#1}{#2}}
  \else%
    \global\@minipagefalse%
    \hbox to\hsize{\box\@tempboxa}
  \fi%
  \addtolength{\hsize}{-\algomargin}%
}
\makeatother

\def\Bka{{\it Biometrika}}
\def\Pro{\mathrm{pr}}
\def\Pb{Poisson-binomial\ }

\begin{document}

\jname{Biometrika}
\copyrightinfo{\Copyright\ 2014 Biometrika Trust\goodbreak {\em Printed in Great Britain}}


\markboth{S. Hidaka}{Miscellanea}

\title{General type-token distribution}

\author{S. Hidaka}
\affil{School of Knowledge Science, Japan Advanced Institute of Science and Technology,\\ 1-1 Asahidai, Nomi, Ishikawa, Japan \email{shhidaka@jaist.ac.jp} }

\maketitle

\begin{abstract}
We consider the problem of estimating the number of types in a corpus using the number of types observed in a sample of tokens from that corpus.
We derive exact and asymptotic distributions for the number of observed types, conditioned upon the number of tokens and the latent type distribution.
We use the asymptotic distributions to derive an estimator of the latent number of types and we validate this estimator numerically.
\end{abstract}


\section{Introduction}
Estimation of the number of unique types or distinct species in a group is required in many fields.
A linguist may study the vocabulary size of an author \citep{Jarvis2002,Malvern2002,Malvern2012,McCarthy2007,McCarthy2010,Tweedie1998,Zipf1949}.
An ecologist may estimate species abundance in a region \citep{Chao1984,Chao1992,Good1953,Huillet2009}.
In such situations, the potential types are unknown a priori.
We derive the asymptotic distribution of the number of observed types in a sample, which may be used to estimate this number of latent types.

Consider a sequence of independent and identically distributed random variables $X_{1}, \ldots, X_{M}$,  where each of these is an integer $X_{i} \in \bar{N} \equiv \left\{ 1, \ldots, N \right\}$ drawn with probability $p_{k} \equiv \Pro (X_{i} = k )$. 
We associate several quantities with this sequence: the number $f_{k, M}$ of integers which appear exactly $k$ times in the sequence, 
the number of tokens $M = \sum_{k=1}^M k f_{k, M}$, 
the number of distinct types $K = \sum_{k=1}^M f_{k, M}$ observed in the sample of $M$ tokens, the latent number of types $N = \sum_{k=0}^M f_{k, M}$, and the word distribution
\[
 \mathcal{N} \equiv  ( p_{1}, \ldots, p_{N} ),\quad p_{i}>0,\quad \sum_{i=1}^{N}p_{i} = 1.
\] 

Past studies have taken two distinct approaches \citep{Bunge1993}.
The first approach utilizes the observation that, if prior samples reflect the probability that a subsequent one is of a given type, then this implies that the frequencies $f_{k, M}$  satisfy certain relations \citep{Good1953,Goodman1949,Ewens1972,Pitman1995}.
Typically, the number of tokens $M$ is fixed.
The second approach is to fit a curve to pairs $( K, M )$ of the number of types $K$ observed in $M$ tokens \citep{Brainerd1982,Chao1992,Herdan1960,Malvern2002,McCarthy2010,Tweedie1998}.
The pairs $(K, M)$ used in this approach are derived from an empirical data set, and the number of latent types is a parameter in the fitting model.

Our work builds upon the second approach by deriving the probability distribution of the pairs $(K, M)$.
This distribution is implicit in \citet{Brainerd1982}, who derived its first- and second-order moments.

\section{Type-token distribution}
\subsection{Exact probability distribution}

Suppose that $M$ tokens are drawn from a corpus with word distribution $\mathcal{N} = ( p_{1}, \ldots, p_{N} )$.
For a subset $s \subseteq \bar{N}$, the probability that a sampled word has a type in $s$ is $\Pro( s ) = \sum_{ i \in s }p_{i}$, with $\Pro( \emptyset ) = 0$. 
By the inclusion-exclusion principle \citep{Allenby2011}, the probability that the types observed in $M$ tokens are precisely those in $s$ is 
\begin{equation}
\label{eq-Psubset}
\Pro( s \mid M, \mathcal{N} ) =
\sum_{ k = 0 }^{ | s | - 1 }
( - 1 )^{ k }
\sum_{ \{ t \subseteq s: |t| = k \} }
\Pro\left(  s \setminus t \right)^{M},
\end{equation}
where $s \setminus t$ denotes these elements of $s$ not in $t$.
Equation (\ref{eq-Psubset}) also follows from the Chapman--Kolmogorov equations \citep{Brainerd1972}
\[
 \Pro( s \mid M, \mathcal{N} ) = \Pro( s \mid M-1, \mathcal{N} )\Pro( s ) + \sum_{ i \in s } \Pro( s \setminus \{ i \} \mid M-1, \mathcal{N} )p_{i}.
\] 

For $K = 1, \ldots, N$, the probability that exactly $K$ types occur in a sample of $M$ tokens is 
\begin{equation}
\label{eq-TypeDistribution}
 \Pro( K \mid M, \mathcal{N} ) = 
\sum_{ \{ u \subseteq \bar{N}: |u| = K \} }\Pro( u \mid M, \mathcal{N} ).
\end{equation}
For each set $s \subseteq \{1, \ldots, N\}$ with $|s| = k \leq K$, upon making the substitutions specified by (\ref{eq-Psubset}), 
the expression $\Pro( s )^M$ occurs $( N - k )! / \{ ( N - K )!( K - k )! \}$ times in (\ref{eq-TypeDistribution}).
Therefore,
\begin{equation}
\Pro( K \mid M, \mathcal{N} ) =
\sum_{k = 1}^{K}
(-1)^{K-k}
{{ N-k }\choose{ N-K }}
\sum_{ \{ s \subseteq \bar{N}: |s| = k \} } \Pro( s )^{M}.
\label{eq-ProbabilityOfTypes} 
\end{equation}
We call $\Pro( K \mid M, \mathcal{N} )$ the type-token distribution.

\subsection{Moment-generating function}
\begin{lemma}
\label{lem-MGF}
The moment-generating function of the type-token distribution (\ref{eq-ProbabilityOfTypes}) 
is 
\[
\mathcal{M}_{P, M}(t) 
=
\sum_{k = 1}^{N}
\sum_{ \{ s \subseteq \bar{N}: |s| = k \} } \Pro( s )^{M}
e^{kt}
( 1 - e^{t} )^{N-k}.
\]
\end{lemma}

\begin{proof}
By (\ref{eq-ProbabilityOfTypes}), 
\begin{eqnarray}
\mathcal{M}_{P, M}(t) &\equiv& \sum_{K=1}^{N}e^{Kt}\Pro( K \mid M, \mathcal{N} ) \nonumber \\
&=& 
\sum_{k = 1}^{N}
\sum_{ \{ s \subseteq \bar{N} : |s| = k \} } \Pro( s )^{M}
\sum_{K=k}^{N} 
(-1)^{K-k}
{{ N-k }\choose{ N-K }}
e^{ K't }
\nonumber
\\
&=& 
\sum_{k = 1}^{N}
\sum_{ \{ s \subseteq \bar{N} : |s| = k \} } \Pro( s )^{M}
e^{ kt }
\sum_{K'= 0}^{N'} 
(-1)^{K'}
{{ N' }\choose{ N'-K' }}
e^{ K't }.
\nonumber
\end{eqnarray}
This yields Lemma~\ref{lem-MGF} as, by the binomial theorem, 
\[
 \sum_{K'= 0}^{N'} (- e^{t} )^{K'} {{ N' }\choose{ N'-K' }} = ( 1 - e^{t} )^{N'}.
\]
\end{proof}

\subsection{Asymptotic distribution \label{sec-Asymptotic} }
Exact calculation of the type-token distribution (\ref{eq-ProbabilityOfTypes}) is intractable when sampling from corpora with large numbers of types.
It is useful to have a reasonable approximation to this distribution which can be computed efficiently.
We show that \Pb distributions \citep{Chen1997,Shah1994,Wang1993} provide such approximations.

\Pb distributions can be computed efficiently \citep{Fernandez2010,Shah1994}.
Le Cam's (1960) theorem\nocite{LeCam1960}, which provides a Poisson approximation to \Pb distributions, can make computation even more efficient at the cost of accuracy.

\begin{theorem}
\label{thm-PoissonBinomial}
For each positive integer $M$ and $i = 1, \ldots, N$, 
write $s_{i} = \{ 1, \ldots, N \} \setminus \{ i \}$ and $q_{M, i} = 1 - \Pro\left( s_{i} \right)^{M}$.
Consider the family of \Pb distributions 
\begin{equation}
Q( K \mid M, \mathcal{N})
= 
\sum_{ \{ s \subseteq \bar{N}: |s| = K \} }\prod_{ i \in s } q_{M, i} \prod_{ j \in \bar{N} \setminus s } (1 - q_{M, j}).
  \label{eq-PoissonBinomial}
\end{equation}
For a fixed probability distribution $\mathcal{N}$, 
\[
 \lim_{M \rightarrow \infty} \max_{ K = 1, \hdots, N }\left| \Pro( K \mid M, \mathcal{N} ) - Q( K \mid M, \mathcal{N} ) \right|  = 0.
\]

\end{theorem}

\begin{proof}
The moment-generating function of $Q( K \mid M, \mathcal{N} )$ is \citep{Wang1993}
\[
 \mathcal{M}_{Q, M}(t)
= \prod_{ i = 1 }^N\left\{ e^{t} + (1 - e^{t})\Pro( s_{i} )^{M} \right\}.
\]

By Lemma~\ref{lem-MGF},
$ \mathcal{M}_{P, M}(t) = \sum_{ k = 0 }^{N}
 e^{t(N-k)}(1 - e^{t})^{k}\sum_{ \{ s \subseteq \bar{N}: |s| = k \} }\Pro( \bar{N} \setminus s )^{M}$.
Writing $\Delta_{s, M} \equiv \Pro( \bar{N} \setminus s )^{M} - \prod_{ i \in s } \Pro( s_{i} )^{M} $, 
\begin{equation}
 \mathcal{M}_{ P, M }(t) - \mathcal{M}_{ Q, M }(t) = 
\sum_{ k = 2 }^{N}
e^{t(N-k)}\left( 1 - e^{t}\right)^{k}\sum_{ \{ s \subseteq \bar{N}: |s| = k \} }\Delta_{s}^{M},
\nonumber
\end{equation}
Since $- \prod_{i \in s}\Pro( s_{i} )^M \leq \Delta_{s, M} \leq 0$, 
and since the number of subsets $s$ is independent of $M$.
$\lim_{ M \rightarrow \infty} \mathcal{M}_{ P, M }(t) - \mathcal{M}_{ Q, M }(t) = 0$.
As the probability distributions $\Pro( K \mid M, \mathcal{N} )$ and $Q( K \mid M, \mathcal{N} )$ have the same support, this proves the theorem.
\end{proof}

\section{Estimation}
Given  $n$ independent pairs of numbers of types and tokens $( K_{i}, M_{i} )$ ($i = 1, \hdots, n$),  the likelihood of the parameter  $\mathcal{N} = ( p_{1}, \ldots, p_{N} )$ is  
\begin{equation}
\label{eq-Likelihood}
 L( \mathcal{N} ) = \prod_{ i = 1 }^{n} Q( K_{i} \mid M_{i}, \mathcal{N}),
\end{equation}
where $Q$ is the \Pb distribution of (\ref{eq-PoissonBinomial}).
We obtain an estimator  $\bar{N}$ for the number of latent types by maximizing the likelihood $L( \mathcal{N} )$.

Suppose that infinitely many  tokens are sampled from the distribution $\nu = \left( \pi_{1}, \ldots, \pi_{N_{\nu}} \right)$ and that, for each positive integer $M$, there are $K( M )$ types observed amongst the first $M$ tokens. 
For $\mathcal{N} = ( p_{1}, \ldots, p_{N} )$ and $ i = 1, \ldots, N$, by the law of large numbers, the proportion of the tokens of $i$ amongst the first $M$ tokens tends to $p_{i}$ as $M \rightarrow \infty$. Therefore, as $M \rightarrow \infty$, $L_{M}( \mathcal{N} )$ tends to 1 if $N = N_{ \nu } $ and to 0 otherwise.
This proves that the maximum likelihood estimator consistently estimates the number of types.

As a consequence, the optimization of the likelihood function (\ref{eq-Likelihood}) may be restricted to any family of distributions in which, for any positive integer $N$, there is at least one distribution with $N$ types.
When analyzing data from a natural corpus, one may restrict the maximization to the family of Zipf distributions.
This is justified by the prevalence of these distributions in such data \citep{Kornai2002,Zipf1949}.

In our analysis, we compared this estimator to the Good--Turing estimator \citep{Good1953,Gale1995} and the Horvitz--Thompson (1952) estimator\nocite{Horvitz1952}.
We observed that the \Pb estimator was less biased than the other estimators.
See the Supplementary Material.

\section*{Acknowledgements}
The author is grateful to Takuma Torii, Akira Masumi and Dr. Neeraj Kashyap for their helpful discussions and comments on early versions of the manuscript. 
This work was supported by the Artificial Intelligence Research Promotion Foundation, JSPS KAKENHI Grant-in-Aid for Scientific Research B and the Grant-in-Aid for Challenging Exploratory Research.

\section{Supplementary material}
Supplementary material available at \Bka  {} online describes practical use of the \Pb estimator and compares it to two other commonly used type estimators.

\bibliographystyle{biometrika}
\bibliography{ReferencesTypeToken}

\begin{thebibliography}{7}
\expandafter\ifx\csname natexlab\endcsname\relax\def\natexlab#1{#1}\fi

\bibitem[{Carroll(1865)}]{Carroll1865}
\textsc{Carroll, L.} (1865).
\newblock Alice's {A}dventures in {W}onderland.
\newblock \url{http://www.gutenberg.org/ebooks/11}.
\newblock Accessed: March 9th, 2010.

\bibitem[{Carroll(1871)}]{Carroll1871}
\textsc{Carroll, L.} (1871).
\newblock Through the {L}ooking-{G}lass.
\newblock \url{http://www.gutenberg.org/ebooks/12}.
\newblock Accessed: March 9th, 2010.

\bibitem[{Dempster et~al.(1977)Dempster, Laird \& Rubin}]{Dempster1977}
\textsc{Dempster, A.~P.}, \textsc{Laird, N.~M.} \& \textsc{Rubin, D.~B.}
  (1977).
\newblock Maximum likelihood from incomplete data via the em algorithm.
\newblock \textit{Journal of Royal Statistical Society Series B} \textbf{39},
  1--38.

\bibitem[{Gale \& Sampson(1995)}]{Gale1995}
\textsc{Gale, W.~A.} \& \textsc{Sampson, G.} (1995).
\newblock Good--{T}uring frequency estimation without tears.
\newblock \textit{Journal of Quantitative Linguistics} \textbf{2}, 217--237.

\bibitem[{Good(1953)}]{Good1953}
\textsc{Good, I.~J.} (1953).
\newblock The population frequencies of species and the estimation of
  population parameters.
\newblock \textit{Biometrika} \textbf{40}, 237--264.

\bibitem[{Horvitz \& Thompson(1952)}]{Horvitz1952}
\textsc{Horvitz, D.~G.} \& \textsc{Thompson, D.~J.} (1952).
\newblock A generalization of sampling without replacement from a finite
  universe.
\newblock \textit{Journal of the American Statistical Association} \textbf{47},
  663--685.

\bibitem[{Zechmeister et~al.(1995)Zechmeister, Chronis, Cull, D'Anna \&
  Healy}]{Zechmeister1995}
\textsc{Zechmeister, E.~B.}, \textsc{Chronis, A.~M.}, \textsc{Cull, W.~L.},
  \textsc{D'Anna, C.~A.} \& \textsc{Healy, N.~A.} (1995).
\newblock Growth of a functionally important lexicon.
\newblock \textit{Journal of Literacy Research} \textbf{27}, 201--212.

\end{thebibliography}


\begin{thebibliography}{27}
\expandafter\ifx\csname natexlab\endcsname\relax\def\natexlab#1{#1}\fi

\bibitem[{Allenby \& Slomson(2011)}]{Allenby2011}
\textsc{Allenby, R.~B.} \& \textsc{Slomson, A.} (2011).
\newblock \textit{How to Count: An Introduction to Combinatorics}.
\newblock Florida, USA: CRC Press.

\bibitem[{Brainerd(1972)}]{Brainerd1972}
\textsc{Brainerd, B.} (1972).
\newblock On the relation between types and tokens in literary text.
\newblock \textit{Journal of Applied Probability} \textbf{9}, pp. 507--518.

\bibitem[{Brainerd(1982)}]{Brainerd1982}
\textsc{Brainerd, B.} (1982).
\newblock On the relation between the type-token and species-area problems.
\newblock \textit{Journal of Applied Probability} \textbf{19}, pp. 785--793.

\bibitem[{Bunge \& Fitzpatrick(1993)}]{Bunge1993}
\textsc{Bunge, J.} \& \textsc{Fitzpatrick, M.} (1993).
\newblock Estimating the number of species: A review.
\newblock \textit{Journal of the American Statistical Association} \textbf{88},
  364--373.

\bibitem[{Chao(1984)}]{Chao1984}
\textsc{Chao, A.} (1984).
\newblock Nonparametric estimation of the number of classes in a population.
\newblock \textit{Scandinavian Journal of Statistics} \textbf{11}, pp.
  265--270.

\bibitem[{Chao(1992)}]{Chao1992}
\textsc{Chao, M.-T.} (1992).
\newblock From {A}nimal {T}rapping to {T}ype-{T}oken.
\newblock \textit{Statistica Sinica} \textbf{2}, 189--201.

\bibitem[{Chen \& Liu(1997)}]{Chen1997}
\textsc{Chen, S.~X.} \& \textsc{Liu, J.~S.} (1997).
\newblock Statistical applications of the {P}oisson-binomial and conditional
  {B}ernoulli distributions.
\newblock \textit{Statistica Sinica} \textbf{7}, 875--892.

\bibitem[{Ewens(1972)}]{Ewens1972}
\textsc{Ewens, W.~J.} (1972).
\newblock The sampling theory of selectively neutral alleles.
\newblock \textit{Theoretical Population Biology} \textbf{3}, 87--112.

\bibitem[{Fernandez \& Williams(2010)}]{Fernandez2010}
\textsc{Fernandez, M.} \& \textsc{Williams, S.} (2010).
\newblock Closed-form expression for the {P}oisson-binomial probability density
  function.
\newblock \textit{IEEE Transactions on Aerospace Electronic Systems}
  \textbf{46}, 803--817.

\bibitem[{Gale \& Sampson(1995)}]{Gale1995}
\textsc{Gale, W.~A.} \& \textsc{Sampson, G.} (1995).
\newblock Good--{T}uring frequency estimation without tears.
\newblock \textit{Journal of Quantitative Linguistics} \textbf{2}, 217--237.

\bibitem[{Good(1953)}]{Good1953}
\textsc{Good, I.~J.} (1953).
\newblock The population frequencies of species and the estimation of
  population parameters.
\newblock \textit{Biometrika} \textbf{40}, 237--264.

\bibitem[{Goodman(1949)}]{Goodman1949}
\textsc{Goodman, L.~A.} (1949).
\newblock On the estimation of the number of classes in a population.
\newblock \textit{Annals of Mathematical Statistics} \textbf{20}, 572--579.

\bibitem[{Herdan(1960)}]{Herdan1960}
\textsc{Herdan, G.} (1960).
\newblock \textit{Type-{T}oken {M}athematics: A {T}extbook of {M}athematical
  {L}inguistics}.
\newblock Hague, Netherlands: Mouton \& Co.

\bibitem[{Horvitz \& Thompson(1952)}]{Horvitz1952}
\textsc{Horvitz, D.~G.} \& \textsc{Thompson, D.~J.} (1952).
\newblock A generalization of sampling without replacement from a finite
  universe.
\newblock \textit{Journal of the American Statistical Association} \textbf{47},
  663--685.

\bibitem[{Huillet \& Paroissin(2009)}]{Huillet2009}
\textsc{Huillet, T.} \& \textsc{Paroissin, C.} (2009).
\newblock Sampling from {D}irichlet partitions: estimating the number of
  species.
\newblock \textit{Environmetrics} \textbf{20}, 853--876.

\bibitem[{Jarvis(2002)}]{Jarvis2002}
\textsc{Jarvis, S.} (2002).
\newblock Short texts, best-fitting curves and new measures of lexical
  diversity.
\newblock \textit{Language Testing} \textbf{19}, 57--84.

\bibitem[{Kornai(2002)}]{Kornai2002}
\textsc{Kornai, A.} (2002).
\newblock How many words are there?
\newblock \textit{Glottometrics} \textbf{4}, 2002.

\bibitem[{Le~Cam(1960)}]{LeCam1960}
\textsc{Le~Cam, L.} (1960).
\newblock An approximation theorem for the {P}oisson binomial distribution.
\newblock \textit{Pacific Journal of Mathematics} \textbf{10}, 1181--1197.

\bibitem[{Malvern \& Richards(2002)}]{Malvern2002}
\textsc{Malvern, D.} \& \textsc{Richards, B.} (2002).
\newblock Investigating accommodation in language proficiency interviews using
  a new measure of lexical diversity.
\newblock \textit{Language Testing} \textbf{19}, 85--104.

\bibitem[{Malvern \& Richards(2012)}]{Malvern2012}
\textsc{Malvern, D.} \& \textsc{Richards, B.} (2012).
\newblock \textit{Measures of Lexical Richness}.
\newblock Oxford, UK: Blackwell Publishing Ltd.

\bibitem[{McCarthy \& Jarvis(2010)}]{McCarthy2010}
\textsc{McCarthy, P.} \& \textsc{Jarvis, S.} (2010).
\newblock {M}{T}{L}{D}, vocd-{D}, and {H}{D}-{D}: A validation study of
  sophisticated approaches to lexical diversity assessment.
\newblock \textit{Behavior Research Methods} \textbf{42}, 381--392.
\newblock 10.3758/BRM.42.2.381.

\bibitem[{McCarthy \& Jarvis(2007)}]{McCarthy2007}
\textsc{McCarthy, P.~M.} \& \textsc{Jarvis, S.} (2007).
\newblock vocd: A theoretical and empirical evaluation.
\newblock \textit{Language Testing} \textbf{24}, 459--488.

\bibitem[{Pitman(1995)}]{Pitman1995}
\textsc{Pitman, J.} (1995).
\newblock Exchangeable and partially exchangeable random partitions.
\newblock \textit{Probability Theory and Related Fields} \textbf{102},
  145--158.

\bibitem[{Shah(1994)}]{Shah1994}
\textsc{Shah, B.~K.} (1994).
\newblock On the distribution of the sum of independent integer valued random
  variables.
\newblock \textit{American Statistician} \textbf{27}, 123--124.

\bibitem[{Tweedie \& Baayen(1998)}]{Tweedie1998}
\textsc{Tweedie, F.} \& \textsc{Baayen, R.} (1998).
\newblock How variable may a constant be? measures of lexical richness in
  perspective.
\newblock \textit{Computers and the Humanities} \textbf{32}, 323--352.
\newblock 10.1023/A:1001749303137.

\bibitem[{Wang(1993)}]{Wang1993}
\textsc{Wang, Y.~H.} (1993).
\newblock On the number of successes in independent trials.
\newblock \textit{Statistica Sinica} \textbf{3}, 295--312.

\bibitem[{Zipf(1949)}]{Zipf1949}
\textsc{Zipf, G.~K.} (1949).
\newblock \textit{Human {B}ehavior and the {P}rinciple of {L}east-{E}ffort}.
\newblock Oxford, UK: Addison-Wesley Press.

\end{thebibliography}

\end{document}


\jname{Biometrika}
\jyear{}
\jvol{}
\jnum{}
\copyrightinfo{\Copyright\ 2012 Biometrika Trust\goodbreak {\em Printed in Great Britain}}


\markboth{S. Hidaka}{Supplementary material}

\title{Supplementary material to General type-token distribution}

\author{S. Hidaka}
\affil{School of Knowledge Science, Japan Advanced Institute of Science and Technology,\\ 1-1 Asahidai, Nomi, Ishikawa, Japan \email{shhidaka@jaist.ac.jp} }

\maketitle

\section{Practical type estimation}
	
Consider the problem of estimating the size of Lewis Carroll's vocabulary when he wrote ``Alice's Adventures in Wonderland''.
The number of tokens we have from this corpus is limited to the 24,168 words which appear in the novel, and there is little hope of adding to this sample.
In practice, one often has to deal with such limitations on sampling.
The conventional method of dealing with this problem is to generate multiple samples from the same data set for use in estimation.
For example, in the case of ``Alice's Adventures in Wonderland'', one would sample data sets $D_{1}, \ldots, D_{n}$ from the text, with each data set $D_{i}$ consisting of $M_{i}$ tokens.
These data sets would not be independent as required by most estimators.
It has been observed empirically, however, that the use of such data sets increases the accuracy of estimators when additional sampling is difficult.

There are many schemes one could use to generate the data sets $D_{1}, \ldots, D_{n}$.
Our objective is to compare type estimators.
We therefore adopt the strategy of sampling successive tokens: if the original sample consists of  $M$ tokens, we decide upon a target number $n \le M$ of data sets and take for $D_{i}$ the first $[M/ n] \times i$ tokens, where $[x]$ denotes the greatest integer less than or equal to $x$.
We take care to choose $n$ so that the overlap between data sets does not impede estimation.

The Good--Turing \citep{Good1953,Gale1995} and Horvitz--Thompson estimators \citep{Horvitz1952} are most commonly used in practice.
These estimators make use of the frequency $f_{k, M}$ defined in the introduction to our article. We denote by $\hat{N}_{\text{GT}}$ the Good--Turing estimate of the latent number of types, and by $\hat{N}_{\text{HT}}$ the Horvitz--Thompson estimate.
These are
\[
 \hat{N}_{\text{GT}} \equiv f_{1, M} + K,\quad 
\hat{N}_{\text{HT}} \equiv \sum_{k=1}^{\infty} \frac{ f_{k, M} }{ 1 - \left( 1 - \frac{k}{M} \right)^{M}}\ .
\]

We compared these estimators to the maximum likelihood estimator for the likelihood function $L( \mathcal{N} )$ of \EqLikelihood.
In maximizing this likelihood, we assumed that $\mathcal{N}$ was a Zipf distribution on the set $\bar{N}$ for some positive integer $N$. This constraint makes the optimization tractable and, as noted in the main article, it does not affect the consistency of the estimator.

The Zipf distributions form a two-parameter family. 
Each distribution $\Pro( k ) \propto k^{-a}, k = 1, \ldots, N$, is specified by its exponent $a$ and the size $N$ of its support.
For such a distribution $\mathcal{N}$, we write $L( \mathcal{N} ) = L( a, N )$.
We obtained maximum likelihood estimates $\hat{a}$ and $\hat{N}$ of these parameters, using $\hat{N}_{\text{PB}} \equiv \hat{N}$ as the \Pb estimate of the latent number of types.

\section{Numerical experiments \label{sec-NumericalExperiments}}

We assessed these estimators using two classes of numerically generated data sets $D$.
The data sets in the first class consisted of $M = 1000, 1500, 2000$ tokens sampled from a corpus of $N = 1000$ types according to the Zipf distribution $ p_{k} \propto k^{-a}, a = 1 $.
The data sets in the second class consisted of $M = 2000$ tokens sampled from a corpus of $N = 1000$ types according to the Zipf distributions $ p_{k} \propto k^{-a}, a = 0, 0.5, 1$.

For each sample $D$, we generated data sets $D_{1}, \ldots, D_{M/50}$ by successively sampling tokens as described above. 
We obtained the estimate $\hat{N}_{\text{PB}}$ for the corpus corresponding to $D$ by maximizing the product of the likelihood functions corresponding to each of the data sets $D_{i}$.

In the family of Zipf distributions, the exponent $a$ is a smooth parameter.
Consequently, it is easy to maximize the conditional likelihood $L( a \mid N )$.
However, as $N$ is a discrete parameter, and this does not translate to easy maximization of $L( a, N )$. In these experiments, we assumed that $N \le 2000$ and performed the optimization on $N$ by brute force.

\begin{figure}
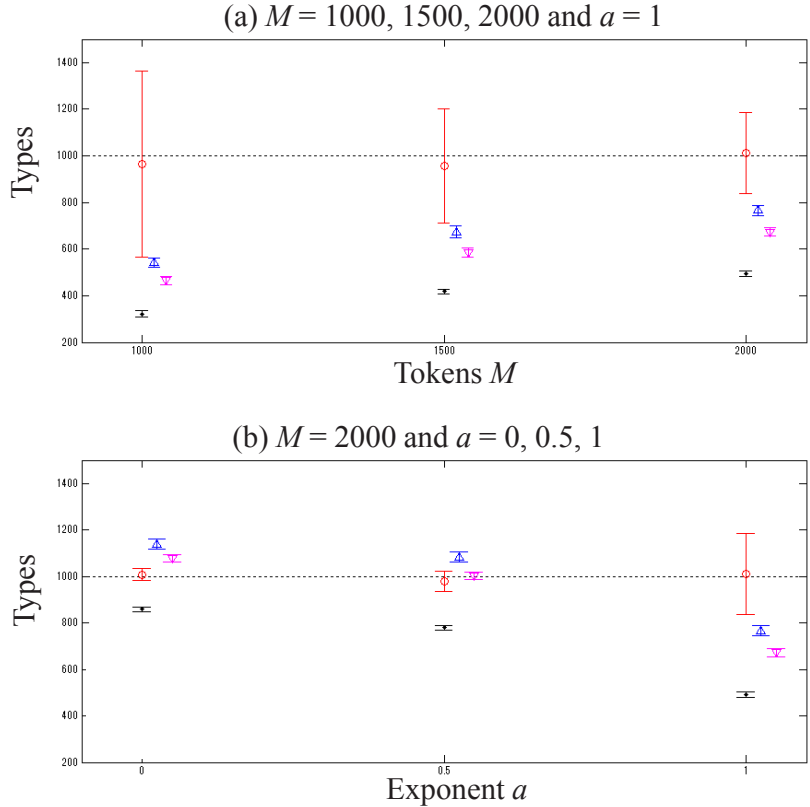

\figurebox{}{\linewidth}{}[SimulatedDataAnalysisWithHT3.eps]
\caption{
Results of simulation study. Panels (a) and (b) show the average estimates produced by the three estimators for each set of parameters.
Red circles represent average \Pb estimates $\hat{N}_{\text{PB}}$; blue, upward-pointing triangles represent average Good--Turing estimates $\hat{N}_{\text{GT}}$; and purple, downward-pointing triangles represent average Horvitz--Thompson estimates $\hat{N}_{\text{HT}}$.
Black dots represent average numbers of observed types for each set of parameters.
The dashed lines reflect the true number of types, $N = 1000$.
The vertical lines around each marker indicate the standard deviation of the estimates on the hundred data sets corresponding to that set of parameters.}
\label{fig-MLE}
\end{figure}

For each choice of parameters $M$ and $a$, we independently generated one hundred data sets $D$ which we used to estimate the size of the underlying corpus. 
The result of this analysis are shown in Fig.~\ref{fig-MLE}.
These results indicate that the Good--Turing and Horvitz--Thompson estimators are more biased for such data than the \Pb estimator.
Moreover, their biases increase with the exponent of the Zipf distributions whereas the mean \Pb estimates consistently reflect the true number of types.

\section{Alice in Wonderland}

We used the Good--Turing, Horvitz--Thompson, and \Pb estimators to estimate the size of Lewis Carroll's vocabulary when he wrote ``Alice's Adventures in Wonderland''\nocite{Carroll1865}.
The text consists of 24,168 words with 4,920 distinct types.
The Good--Turing and Horvitz--Thompson estimates for the size of the underlying corpus were, respectively, 
\[
 \hat{N}_{\text{GT}} = 8346,\quad \hat{N}_{\text{HT}} = 6988.8.
\]
To put this in context, the vocabulary of an average adult native English speaker has been estimated to contain over 20,000 words \citep{Zechmeister1995}.
Taken together, ``Alice's Adventures in Wonderland'' and ``Through the Looking Glass'' \citep{Carroll1865,Carroll1871} contain 8869 distinct words, already exceeding these estimates. 

We derived the \Pb estimate by successively sampling $ n = 48$ data sets $D_{1}, \hdots, D_{48}$ from the text and maximizing the product of corresponding likelihoods.
In this case, we did not find it appropriate to set a hard bound on the number of types and used the brute force approach of the previous section.
As the difficulty of optimization stems from the discrete nature of the parameter $N$ for the family of Zipf distributions, we introduced a smooth parameter $\lambda$ which determines $N$.
We did this by assuming that $N$ is a Poisson random variable with parameter $\lambda$, so that
\[
 \Pro\left( N = k \mid \lambda \right) = \frac{\lambda^{k}}{k!} e^{-\lambda},\quad \lambda > 0,\quad k = 0, 1, \ldots.
\]
Under this assumption, we write $L( \mathcal{N} ) = L( a, \lambda )$.
We use the expectation-maximization algorithm \citep{Dempster1977} to maximize $L( a, \lambda )$.
Given the maximum likelihood estimate $\hat{\lambda}$, the estsimate for the latent number of types was the expected values of the corresponding Poisson random variable, $\hat{N}_{\text{PB}} \equiv \hat{\lambda}$.

The \Pb estimate of the size of Lewis Carroll's vocabulary when he wrote ``Alice's Adventures in Wonderland'' is $\hat{N}_{\text{PB}} = 41,647.128$ (its standard error $191.748$). 

\bibliographystyle{biometrika}
\bibliography{ReferencesTypeToken}